\documentclass[12pt]{article}

\usepackage{amssymb,commath}
\usepackage{epsfig,amssymb,amsfonts,amsmath,graphicx,dsfont,cite,xfrac}
\usepackage{authblk}

\usepackage{cite}
\usepackage[colorlinks=true,linkcolor=blue,citecolor=red]{hyperref}

\newcommand{\be}{\begin{equation}}
\newcommand{\ee}{\end{equation}}
\newcommand{\bea}{\begin{eqnarray}}
\newcommand{\eea}{\end{eqnarray}}

\title{On the geometry of spatial biphoton correlation in spontaneous parametric down-conversion}

\author[1]{Lorenzo M.~Procopio}
\author[1]{Oscar Rosas-Ortiz}
\author[2]{V\'{\i}ctor Vel\'asquez}

\affil[1]{\footnotesize Physics Department, Cinvestav, AP 14-740, 07000
M\'exico DF, Mexico}

\affil[1]{\footnotesize Facultad de Ciencias, UNAM, A.P. 20-364, M\'exico D.F. 04510, Mexico}

\date{}
\begin{document}

\maketitle

\begin{abstract} Analytical expressions are derived for the distribution rates of spatial coincidences in the counting of photons produced by spontaneous parametric down conversion (SPDC). Gaussian profiles are assumed for the wave function of the idler and signal light created in type-I SPDC. The distribution rates describe ellipses on the detection planes that are oriented at different angles according to the photon coincidences in either horizontal-horizontal, vertical-vertical, horizontal-vertical or vertical-horizontal position variables. The predictions are in agreement with the experimental data obtained with a type-I BBO crystal that is illuminated by a 100~mW violet pump laser as well as with the results obtained from the geometry defined by the phase-matching conditions.
\end{abstract}

\section{Introduction}

Nonlinear effects in optics offer the possibility of generating light in almost any manner, they are mostly achieved via the interaction of light with matter \cite{Kly88,Sal91,Man95,Lam07,Men07,Han07,Ou07,Wal10}. The {\em spontaneous parametric down conversion} (SPDC), for example, occurs when a nonlinear crystal is illuminated by the appropriate light. In the crystal one of the incoming photons, at the pump frequency $\omega_p$, is spontaneously annihilated at the time that two new outgoing photons are created at the frequencies $\omega_s$ and $\omega_i$. It is common to name the new photons as {\em signal} and {\em idler} respectively. The polarization properties of the generated pair define the resulting spatial distribution and serve to characterize the SPDC phenomenon. If the polarization of the new pair of photons is parallel to each other and orthogonal to the polarization of the pump photon then the spatial distribution of the created light forms a cone that is aligned with the pump beam propagation and the apex of which is in the crystal. This last condition defines the {\em type-I} SPDC. On the other hand, the light created by the crystal in the {\em type-II} SPDC forms two (not necessarily collinear) cones that are oriented along the propagation of the pump beam and share the same apex located somewhere in the crystal. This last is because in type-II SPDC, the polarization of the idler photons is orthogonal to the polarization of the signal ones.

\begin{figure}[h]
\centering
\includegraphics[scale=0.25]{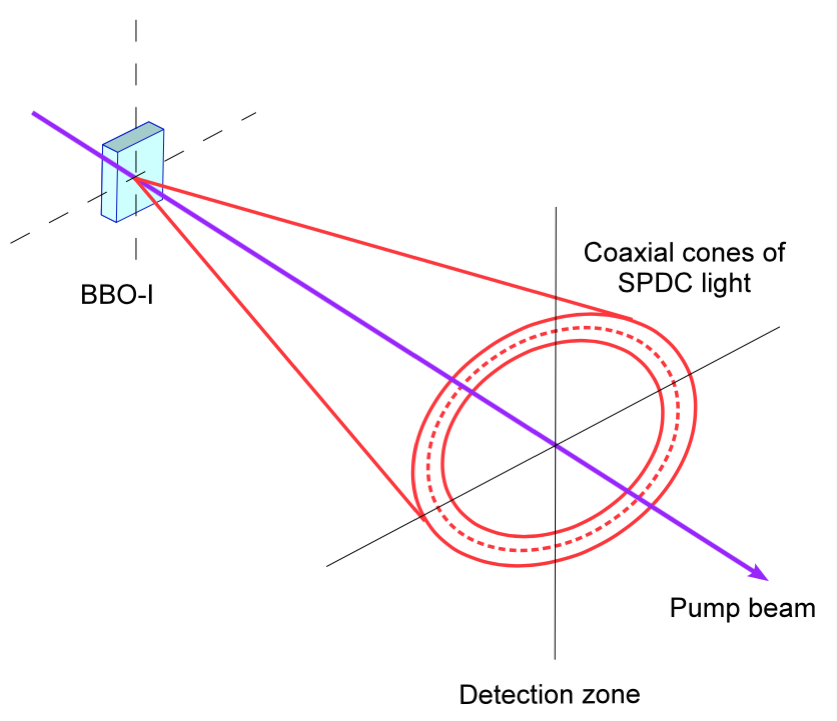}  
\caption{\label{fig1} \footnotesize
Schematic representation of the degenerated spontaneous parametric down-conversion presented in a BBO-I crystal that is pumped by a violet laser diode ($\lambda_p \approx 405.38$ nm). This gives rise to the creation of infra-red photons ($\lambda_i = \lambda_s \approx 810.76$ nm) that form coaxial cones of light. The plane that defines the detection zone is orthogonal to the propagation direction of the pump beam and cuts the cones by forming concentric circles centered at the pump beam. The dotted red circle in the figure represents the ideal detection zone of the infra-red photons created by the annihilation of a violet photon with $\lambda_p = \lambda_0$. That is, once an incoming photon with ideal $\lambda_0$ has been annihilated, each of the two resulting infrared photons will be found with certainty at a place on the dotted circle, one in the antipodal position of the other. The same can be said for the circles associated to the annihilation of incoming photons with wavelength that is close to the ideal one $\lambda_p \approx \lambda_0$.}
\end{figure}

Signal photons produced in SPDC are useful whenever their `passive' partners are detected (i.e., signal photons are triggered by the idler photons). Thus, the signal mode does not exist unless the idler-trigger photon is registered in the detector. In this form, the state of the entire {\em biphoton} system (idler + signal) is conditionally collapsed into a single photon state (in the signal mode) by a detection event in the idler channel. The properties of the signal mode depend on the optical mode of the pump photon and are determined by the way in which the measurement in the idler channel is performed. This non-local preparation of single photons was suggested and tested in the eighties by Hong and Mandel \cite{Hon85,Hon86} and by Grangier, Roger and Aspect \cite{Gra86}, though the spatial correlation of the involved biphoton state was suggested in 1969 by Klyshko \cite{Kly69} and reported in 1970 by Burnham and Weinberg \cite{Bur70}. It is important to stress that the temporal characteristics of the single signal photon pulses are not controlled since the pair production is spontaneous. Moreover, in the SPDC process the efficiency of the photon pair production is very low ($\approx 10^7 -10^{11}$). Thereby pumped nonlinear crystals seem to be not the best sources of single photons \cite{Lou05} (see also \cite{Lam07}). However, since its introduction in 1986, the conditional single photon production is a fundamental ingredient in many of the contemporary experiments of quantum control and quantum information  \cite{Men07,Bou00,Mie09,Zei10}. Indeed, the non-local preparation of single photons means that the idler photon is correlated to the signal one in at least one (and the same) of the variables that define their quantum states. If such a correlation is preserved after the non-destructive manipulation of the involved variable(s) one says that the idler and signal photons are entangled (see e.g. \cite{Mie09,Acz03} and references quoted therein). Thus, the study of correlations between the variables of multipartite systems is meaningful in physics from both theoretical and experimental approaches. The SPDC is in this sense useful since biphoton states can be constructed such that they are correlated in their spatial, temporal, spectral or polarization properties (see, e.g. \cite{Men07}, Ch. 4.4.5, and references quoted therein). In previous works \cite{Pro09,Pro10} we have reported the observation of spatial correlations between the photons created by a (type-I) nonlinear uniaxial crystal of Beta-barium borate ($\beta$--$B_aB_2O_4$, BBO-I for simplicity) that is pumped by a 100 mW violet laser diode operating at 405.38 nm and bandwidth 0.78 nm. There, we developed a simple geometric model to predict the distribution rates of spatial coincidences in the idler and signal channels when the photon collectors are displaced in either horizontal or vertical directions in a plane (detection zone) that is transversal to the pump beam and is located a distance $L$ from the center of mass of the crystal. The geometric model is based on the conservation of energy
\be
\omega_p = \omega_i + \omega_s,
\label{energy}
\ee
and momentum 
\be
\mathbf k_p = \mathbf k_s + \mathbf k_i.
\label{momentum}
\ee
This last expression is known as {\em phase-matching} condition and implies that the $\mathbf k_r$, with $r=p,i,s$, are all coplanar. In the degenerate case the new photons have the same frequency $\omega_i = \omega_s = \omega_p/2$, so that $\lambda_i =\lambda_s$ (actually, 
$\omega_i \approx \omega_s$ and $\lambda_i  \approx\lambda_s$). If the idler photon is detected along a particular direction $\mathbf k_i$, then its signal partner is along the direction defined by $\mathbf k_s = \mathbf k_p - \mathbf k_i$. Then, the process is axial symmetric and the SPDC photons emerge from the crystal by forming cones that produce a ring pattern on the detection zone. To be explicit, if $\lambda_0$ is the ideal wavelength leading to the type-I SPDC in a given nonlinear crystal, then the $\lambda_0$-incoming light which is spontaneously annihilated gives rise to the creation of light with twice the incoming wavelength that propagates from the crystal towards the detection zone by forming a cone. A transversal cut of such a cone defines a circle centered at the pump beam (this is represented by the dotted circle depicted in Figure~\ref{fig1}). Other incoming photons with $\lambda_p \approx \lambda_0$ will produce cones coaxial to the one associated to $\lambda_0$ and will define a ring of width $2\delta$ that is bounded by an inner circle of diameter $d-\delta$, and an external circle of diameter $d+\delta$, with $d$ the diameter of the ideal circle. Given $\delta \geq 0$, the created photon pairs are emitted on opposite sides of the corresponding cone (antipodal points on the related circle in the detection ring). The parameter $\delta$ represents the arbitrariness in the wavelength (equivalently, the frequency) of the incoming photons that is associated to actual light sources as well as the approximate validity of the conservation laws (\ref{energy}) and (\ref{momentum}) in the lab. Assuming that the parameter $\delta$ is smaller than the diameter $d$ one can calculate the rate of coincidences in the simultaneous detection of idler and signal photons on the ring. The result is in agreement with the measurements within the experimental accuracy (full details in \cite{Pro09}). An issue is whether the aforementioned geometric model can be placed on quantum theoretical grounds.

In this work we address the problem of determining the approximations that are necessary to reproduce the results reported in \cite{Pro09,Pro10} from a formal quantum approach. The framework is suitable since our geometric model has been successfully verified in the lab. Theoretical treatments of SPDC that include the spatial correlations between idler and signal  photons as a main subject can be found in e.g.,  \cite{Kly88,Man95,Ou07,Wal10,Hon85,Kly69}. Our approach is guided by the fact that the ring in the detection zone has a finite width $2\delta$, with $\delta$ determined by the general properties of the pump beam like the spatial profile, beam and spectral widths, wavelength, and so on. Taking this into account the mean and variance of the set of spatial coincidences are expected to be finite (a fact that has been verified in the lab). Therefore, assuming that the counts are large enough, the central limit theorem indicates that the rates of coincidences should be normal (Gaussian) distributions of the position. Indeed, the geometric model assumes Gaussian profiles for the (idler and signal) horizontal spatial distributions determined by the right-handed Cartesian system of coordinates that has its origin at the center of the ring and the $z$-axis of which is along the direction of propagation of the pump beam (see Figure~\ref{fig1}). References \cite{Joo96,Mol03,Tor04,Mol05} deal with approaches that are close to the one we are going to present here. The organization of the paper is as follows. In Section~\ref{spdc} we review the generalities of the SPDC phenomenon. First, the Hamiltonian of interaction between the pump light and the nonlinear crystal is constructed (Section~\ref{hamiltoniano}). Then the corresponding time-evolution operator is applied to the quantum state of the three fields (pump, idler and signal) in order to get the probability amplitudes that serve to calculate the spatial correlations between the idler and signal photons (Section~\ref{time}). In Section~\ref{geometria} we make several simplifying assumptions and approximations to recover the results of the geometric model (Section~\ref{aprox}). Some of our main results are discussed in Section~\ref{main}. Concluding remarks are given at the very end of the paper.

\section{Spontaneous parametric down-conversion revisited}
\label{spdc}

\subsection{The Hamiltonian}
\label{hamiltoniano}

Let us express the classical optical electric field as the sum of its positive and negative frequency parts:
\begin{equation}
\mathbf E (\mathbf{r},t)=\mathbf{E}^{(+)}(\mathbf{r},t) + \mathbf{E}^{(-)} (\mathbf{r},t),
\label{cfield1}
\end{equation}
where
\begin{equation}
\mathbf{E}^{(+)} (\mathbf{r},t)= \frac{1}{\sqrt{V}} \displaystyle\sum_{\mathbf k, \nu} i (2\pi \hbar \omega)^{1/2}   \mathbf{e}_{\mathbf{k},\nu}  \alpha_{\mathbf{k},\nu} \exp [i(\mathbf{k}\cdot\mathbf{r}-\omega t)] = [\mathbf{E}^{(-)} (\mathbf{r},t)]^*.
\label{cfield2}
\end{equation}
Here $z^*$ stands for the complex conjugate of $z \in \mathbb C$ while $\mathbf{e}_{\mathbf{k},\nu}$, $\mathbf k$, $\omega$ and $\alpha_{\mathbf{k},\nu}$ represent the two-dimensional polarization vector, the wave vector, the frequency and the mode amplitude of the field respectively. The volume $V$ has been introduced for quantization and the indices $\nu$ and $\mathbf k$ are summed over 2 and all possible wave vectors respectively. 

The field represented by each of the parts of $\mathbf E (\mathbf{r},t)$ in (\ref{cfield1}) is intrinsically complex. According to Glauber, the quantized version of these variables, namely $\hat{\mathbf E}^{(+)} (\mathbf{r},t)$ and $\hat{\mathbf E}^{(-)} (\mathbf{r},t)$, act to change the state of the field in different ways, the former associated to photon absorption and the latter to photon emission \cite{Gla07}, Ch.1 pp. 1-21. The state in which the field is empty of all photons, denoted $\vert vac \rangle$, can be defined as the solution of the equation
\be
\hat{\mathbf E}^{(+)} (\mathbf{r},t) \vert vac \rangle = 0,
\label{vac}
\ee
with adjoint relation
\be
\langle vac  \vert \hat{\mathbf E}^{(-)} (\mathbf{r},t)  = 0.
\label{vacr}
\ee
Thus, the frequency positive part operator $\hat{\mathbf E}^{(+)} (\mathbf{r},t)$ is a photon annihilation operator that produces an $(n-1)$-photon state when it acts on an $n$-photon state; the regression ends with the (ket) state $\vert vac \rangle$. The Hermitian adjoint of the annihilation operator, $ \left[ \hat{\mathbf E}^{(+)} (\mathbf{r},t)  \right]^{\dagger} =
\hat{\mathbf E}^{(-)} (\mathbf{r},t)$, acts in reversed order: applied to an $n$-photon state it produces an $(n+1)$-photon state. Remark that the action of these operators is right-handed, that is $\hat{\mathbf E}^{(+)} (\mathbf{r},t)$ and $\hat{\mathbf E}^{(-)} (\mathbf{r},t)$ are annihilation and creation photon operators whenever they act on the right of an $n$-photon (ket) state. On the other hand, if they are applied on the left to the $n$-photon (bra) states, then their roles are interchanged, as it is shown in Equation~(\ref{vacr}) where $\hat{\mathbf E}^{(-)} (\mathbf{r},t)$ annihilates the bra $\langle vac \vert$. The SPDC is necessarily a quantum phenomenon, so that the photo-detection theory pioneered by Glauber \cite{Gla07} is fundamental in our description. Incoming photons are absorbed by the nonlinear crystal at the time that pairs of new photons are emitted by it. The canonical quantization of the field (\ref{cfield1}) can be summarized by making the transformation $\alpha_{\mathbf k, \nu} \rightarrow \hat a_{\mathbf k, \nu}$ in (\ref{cfield2}), with $ \hat a_{\mathbf k, \nu}$ the photon annihilation operator related with the mode $(\mathbf k, \nu)$. This last and its Hermitian adjoint $\hat a_{\mathbf{k},\nu}^{\dagger}$ satisfy the commutation rule
\begin{equation}
[\hat a_{\mathbf{k},\nu},\hat a_{\mathbf{k}',\nu'}^{\dagger}]=\delta_{\nu \nu'} \delta(\mathbf{k}-\mathbf{k}').
\end{equation}
The energy of the entire radiation field is represented by a Hamiltonian operator that includes a free field $H_0$ and an interaction $H_I$ parts. The former is proportional to the number operator $\hat N_{\mathbf k, \nu}$ and adds a global phase to the time-evolution of the field quantum state, therefore we can omit it from the calculations (of course, given a polarization $\nu$, this kind of terms are relevant in photo-detection because the photon-collectors measure average values of the product $E_{\nu}^{(-)}  E_{\nu}^{(+)} \propto a_{\mathbf{k},\nu}^{\dagger} a_{\mathbf{k},\nu} = \hat N_{\mathbf k, \nu}$, but a global phase depending on the eigenvalues of $\hat N_{\mathbf k, \nu}$ makes no changes in such averages). The interaction Hamiltonian $H_I$, on the other hand, may be shown \cite{Wal10} to be of the form (Einstein summation is assumed):
\begin{equation}
 \hat H_I= \frac{1}{2}\int_v d^3r \int_0^{\infty} dt_1 \int_0^{\infty} \! dt_2 \, \chi^{(2)}_{ijk}(t-t_1,t-t_2)\hat E_i(\mathbf{r},t)\hat E_j(\mathbf{r},t_1) \hat E_k(\mathbf{r},t_2),
\label{hi}
\end{equation}
with $\chi^{(2)}_{ijk}$ the components of the second order nonlinear susceptibility tensor
\begin{equation} 
\mathbf{P}=\chi^{(1)} \mathbf{E} + \chi^{(2)}\mathbf{EE} + \chi^{(3)}\mathbf{EEE}+ \cdots
\end{equation}
The integrand in (\ref{hi}) is the result of the product $(\chi^{(2)}\mathbf{EE})  \mathbf E$, so that $\hat E_i$ and $\hat E_j$ correspond to the components of the field that is created by the nonlinear crystal while $\hat E_k$ would represent the components of the pump beam. The introduction of the quantized version of (\ref{cfield1})  into Equation~(\ref{hi}) yields
\begin{eqnarray}
 \hat H_I &=& \frac{1}{2}\int_v d^3r \int_0^{\infty} dt_1 \int_0^{\infty}dt_2 \chi^{(2)}_{ijk} \left[ \hat E_i^{(-)} \hat E_j^{(-)} \hat E_k^{(-)} + \hat E_i^{(-)} \hat E_j^{(-)} \hat E_k^{(+)} 
 \right.
\nonumber
\\[1ex]
&&  \hbox{\hskip1cm}+  \hat E_i^{(-)} \hat E_j^{(+)} \hat E_k^{(-)} + \hat E_i^{(-)} \hat E_j^{(+)} \hat E_k^{(+)} + \hat E_i^{(+)} \hat E_j^{(-)} \hat E_k^{(-)} + \hat E_i^{(+)} \hat E_j^{(-)} \hat E_k^{(+)} 
\nonumber
\\[1ex]
&&
\left .
 \hbox{\hskip2cm}+ \hat E_i^{(+)} \hat E_j^{(+)} \hat E_k^{(-)}+  \hat E_i^{(+)} \hat E_j^{(+)} \hat E_k^{(+)} \right].
\label{hi8}
\end{eqnarray}
The conservation of the energy (as well as the photo-detection criteria discussed above)  is fulfilled by the second and seventh terms in the integrand of this last expression, so that we arrive at the Hamiltonian
\begin{equation}
\hat H_I = \int_{\cal V} d^3 \!r  \, \chi_{ijk}^{(2)} \,  \hat E_i^{(+)}(\mathbf{r},t) \hat E_j^{(+)}(\mathbf{r},t) \hat E_k^{(-)}(\mathbf{r},t) + \mbox{H.c.},
\label{hdef}
\end{equation}
where H.c. stands for Hermitian conjugate, $\cal V$ is the volume of the crystal, and the time-dependence of the field factors has been redefined so that
\begin{equation}
 \chi_{ijk}^{(2)}(\omega',\omega'')= \int_0^{\infty} dt' \int_0^{\infty} dt'' \chi^{(2)}_{ijk}(t',t'') e^{-i(\omega't' + \omega''t'')},
\end{equation}
with $t'=t-t_1$ and $t''=t-t_2$. Using the quantized version of (\ref{cfield2}) and the replacement $\sum_{\mathbf{k}} \rightarrow \frac{\cal V}{(2\pi)^3} \int d^3k$, we obtain
\be
\begin{array}{rl}
\hat H_I =  \!\!\! \displaystyle\int_{\cal V} \!d^3r \!\! \int \! \!d^3k_p \!\! \int \! \!d^3k_s \!\! \int \! \!d^3k_i \!\!\sum_{\nu_p,\nu_s,\nu_i} \chi_{ijk}^{(2)}(\omega_p,\omega_s,\omega_i)(\mathbf{e}_{\mathbf{k_p},\nu_p})_i (\mathbf{e}_{\mathbf{k_s},\nu_p})_j^* (\mathbf{e}_{\mathbf{k_i},\nu_i})_k^* \\[4ex]
\times \hat a_{\nu_p}(\mathbf{k_p}) \hat a_{\nu_s}^{\dagger}(\mathbf{k_s})  \hat a_{\nu_i}^{\dagger}(\mathbf{k_i})   e^{-i (\omega_s+\omega_i-\omega_p)t} 
e^{i\Delta\mathbf{k} \cdot \mathbf{r}} + \mbox{H.c.},
\end{array}
\label{hproc}
\ee
where $\Delta\mathbf{k}= \mathbf{k}_p-\mathbf{k}_s -\mathbf{k}_i $, and the nonlinear susceptibility has been rewritten as
\begin{equation}
\chi_{ijk}^{(2)}(\omega_p,\omega_s,\omega_i)=-i \frac{\cal V}{(2\pi)^9} (2\pi \omega_p)^{1/2} (2\pi \omega_s)^{1/2} (2\pi \omega_i)^{1/2} \chi_{ijk}^{(2)}(\omega',\omega'').
\label{sus}
\end{equation}
In the above expressions we have added the labels $p,s$, and $i$, to distinguish the involved fields as {\em pump}, {\em signal}, and {\em idler} respectively. Considering fields with a definite polarization we can drop the $\nu$-indices (and the related sums as well) from Equation~(\ref{hproc}). If the nonlinear susceptibility ({\ref{sus}) is a frequency slowly-varying function such that it can be considered a constant, we write $ \chi= \chi_{ijk}(\omega_p,\omega_s,\omega_i)(\mathbf{e}_{\mathbf{k_p},\nu_p})_i (\mathbf{e}_{\mathbf{k_s},\nu_p})_j^* (\mathbf{e}_{\mathbf{k_i},\nu_i})_k^* $ to get
\be
\hat H_I = \chi \int_{\cal V}d^3r \int d^3k_p\int d^3k_s  \int \! d^3k_i \, \hat a_p(\mathbf{k_p}) \hat a^{\dagger}_s (\mathbf{k_s})  \hat a^{\dagger}_i (\mathbf{k_i}) e^{i\Delta\mathbf{k} \cdot \mathbf{r}}  e^{-i (\omega_s+\omega_i-\omega_p)t}    +\mbox{H.c.}
\label{hfin}
\ee

\subsection{Time-evolution of the biphoton quantum state}
\label{time}

At any time $t \geq t_0$, with $t_0$ the initial time, the system of interest ${\cal S}$ is a composite of three fields ${\cal S} = {\cal S}_p + {\cal S}_i + {\cal S}_p$, each one of them in a photon-state $\vert \psi_r \rangle$ defined as an element of the Hilbert space ${\cal H}_r$, $r=p, s, i$. Let ${\cal H}_r = \mbox{Span} \{ \vert n_r \rangle \}_{n_r=0}^{+\infty}$ be the representation of ${\cal H}_r$ in the Fock basis $\vert n_r \rangle$, $n_r=0,1,\ldots$ Any state of ${\cal S}_r$ can be written as the superposition
\be
\vert \psi_r \rangle = \sum_{n_r=0}^{+\infty} c_{n_r} \vert n_r \rangle, \quad c_{n_r} \in \mathbb C.
\ee
An arbitrary state of the entire system is written as a linear combination of the Kronecker products between the basis elements of the components as follows (see \cite{Enr13} for a recent review on the Kronecker product):
\be
\vert \psi(t) \rangle = \sum_{n_p, n_s, n_i =0}^{+\infty} c_{n_p, n_s, n_i}(t) \vert n_p \rangle \otimes \vert n_s \rangle \otimes \vert n_i \rangle,
\label{state}
\ee
where the Fourier coefficients $c_{n_p, n_s, n_i}(t)$ are analytic functions of the time $t$. That is, the Hilbert space ${\cal H}$  associated to the entire system ${\cal S}$ is 
\be
{\cal H} = \mbox{Span} \{ \vert n_p \rangle \otimes \vert n_s \rangle \otimes \vert n_i \rangle \}_{n_p, n_s, n_i =0}^{+\infty}.
\label{hspace}
\ee
Given the initial state $\vert \psi(t_0) \rangle$, the time-evolution of the tripartite system ${\cal S}$ is determined by the Hamiltonian (\ref{hfin}) according to the rule
\begin{equation} 
|\psi(t)\rangle \triangleq \hat U(t,t_0) |\psi (t_0) \rangle,
\label{psit}
\end{equation}
where
\begin{equation}
\hat U(t,t_0) \triangleq \mbox{exp}\left[{\frac{1}{i\hbar} \int_{t_0}^t \hat H_i(\tau)d\tau}\right]
\label{top}
\end{equation}
represents the time-ordering operator associated to $\hat H_i(\tau)$. Using (\ref{hfin}) in (\ref{top}) we have
\begin{equation} 
\hat U(t)= \exp \left[ \xi \circ \hat a_p \hat a^{\dagger}_s  \hat a^{\dagger}_i + {\mbox{H.c.}} \right],
\label{u}
\end{equation}
with
\begin{equation}
 \xi =\chi  \int d^3k_p\int d^3k_s  \int d^3k_i   \int_v d^3r e^{i\Delta\mathbf{k} \cdot \mathbf{r}}  \int_{t_0}^t  d\tau  e^{i  (\omega_s+\omega_i-\omega_p)\tau}.
 \label{jiji}
\end{equation}
Here $\xi \circ \hat a_p \hat a^{\dagger}_s  \hat a^{\dagger}_i$ has been introduced to simplify the calculations, this means that the integrals of (\ref{jiji}) embrace the product $\hat a_p \hat a^{\dagger}_s  \hat a^{\dagger}_i$  in their integrand. On the other hand, this last expression uses simplified notation to represent the tripartite Kronecker product $\hat a_p \otimes \hat a^{\dagger}_s  \otimes \hat a^{\dagger}_i$. The latter is an operator defined to act on the elements of the Hilbert space of the entire system (\ref{hspace}). The operator $\hat a_s$, defined to act on ${\cal H}_s$, may be promoted to act on ${\cal H}$ by using the transformation ${\cal H}_s \rightarrow \mathbb I_p \otimes {\cal H}_s \otimes \mathbb I_i$, with $\mathbb I_r$ the identity operator in ${\cal H}_r$. The same can be said for $\hat a_p$, $\hat a_i$, and any other operator $\hat A_r$ acting on ${\cal H}_r$ \cite{Enr13}. The power expansion of (\ref{u}) can be associated to the series\footnote{The conditions that are necessary to substitute (\ref{u}) by the Taylor series (\ref{expand}) within an acceptable error in calculations have been reported in Lorenzo M. Procopio, {\em Spatial distributions of correlated photon pairs generated by spontaneous parametric down-conversion}, Ph.D. Thesis, Physics Department, Cinvestav, Mexico D.F., Mexico, 2014.}:
\begin{equation} 
\hat U(t) = \mathbb I + \xi \circ \hat a_p \hat a^{\dagger}_s  \hat a^{\dagger}_i  +  \left[ \xi \circ \hat a_p \hat a^{\dagger}_s  \hat a^{\dagger}_i \right]^2+ \cdots 
\label{expand}
\end{equation}
where $\mathbb I$ is the identity operator in ${\cal H}$. We are interested in the event where a single pump photon is annihilated and two new photons (idler and signal) are created in the crystal at the time $t$. Thus, we assume that the probability of the simultaneous annihilation of two pump photons plus the creation of four new photons (two idler and two signal) is negligible as compared with the probability of a single SPDC event (in the lab, this can be ensured by using pump lasers of the appropriate intensity. Indeed, the counts of double creation of biphoton states --i.e., double coincidences within the spatial sensitivity of the detectors-- may verify such a statement). Therefore, at a first approach one gets
\begin{equation} 
\hat U(t) \approx \mathbb  I + \xi \circ \hat a_p \hat a^{\dagger}_s  \hat a^{\dagger}_i.
\label{corte}
\end{equation}
At $t_0$ we consider that the pump field is in a Glauber state $\vert \alpha \rangle$ \cite{Gla07}, while the idler and signal channels are empty of all photons. To be specific,
\be
\vert \psi_p (t_0)\rangle = \vert \alpha \rangle, \quad \hat a_p \vert \alpha \rangle  = \alpha \vert \alpha \rangle, \quad \alpha \in \mathbb C,
\label{pin}
\ee
and
\be
\vert \Psi_{SPDC} (t_0) \rangle =\vert vac \rangle \otimes \vert vac \rangle: = \vert 0_s, 0_i \rangle.
\ee
The complex number that labels the coherent state (\ref{pin}) is related to the spectral and spatial amplitudes of the pump beam through the eigenvalue of the optical electric field (\ref{cfield1}), so that this is a function of the wave vector $\mathbf k_p$ and the frequency $\omega_p$ of the pump field. In this form, the initial state of the tripartite system is given by
\be
\vert \psi(t_0) \rangle = \vert \alpha (\mathbf k_p, \omega_p )\rangle \otimes \vert \Psi_{SPDC} (t_0) \rangle.
\label{init}
\ee
Using (\ref{init}) and (\ref{corte}) in (\ref{psit}) we obtain
\be
\vert \psi(t) \rangle  = \left(\mathbb  I + \xi \circ \hat a_p \hat a^{\dagger}_s  \hat a^{\dagger}_i\right) \vert \alpha (\mathbf k_p, \omega_p )\rangle \otimes \vert \Psi_{SPDC} (t_0) \rangle =  \vert\alpha(\mathbf{k_p},\omega_p)\rangle \otimes \vert \Psi_{SPDC} (t)\rangle,
\ee
with
\begin{equation}
\vert \Psi_{\tiny\mbox{SPDC}}\rangle = \vert 0_s, 0_i\rangle + \chi \int d^3k_s  \int d^3k_i \Phi(\mathbf{k_i},\omega_i,\mathbf{k_s}, \omega_s, t) \vert 1_{\mathbf{k_s}},1_{\mathbf{k_s}}\rangle  
\end{equation}
the biphoton quantum state created by the nonlinear crystal at time $t$, and $\vert 1_{\mathbf{k_s}},1_{\mathbf{k_i}}\rangle = \hat a^{\dagger}_s(\mathbf{k_s})  \hat a^{\dagger}_i(\mathbf{k_i}) \vert0_s, 0_i\rangle$ the corresponding two photons Fock state. Relevant information concerning spatial and spectral correlations between the idler and signal photon states is encoded in the $\Phi$-function
\begin{equation}
 \Phi(\mathbf{k_i},\omega_i,\mathbf{k_s}, \omega_s)= \int d^3k_p \alpha(\mathbf{k_p},\omega_p)   \int_{\cal V} d^3r e^{i\Delta\mathbf{k} \cdot \mathbf{r}}  \int_{t_0}^t  d\tau  e^{i  (\omega_s+\omega_i-\omega_p)\tau}.
\label{Phi}
\end{equation}

\section{The geometry of spatial biphoton correlation in SPDC }
\label{geometria}

\subsection{Approximations}
\label{aprox}

Usually the dimensions of the nonlinear crystal are larger than the wavelength of the pump beam, so that we can consider $\nu \rightarrow \infty$ to make the identification
\begin{equation} 
\int_{\cal V} d^3r e^{i\Delta\mathbf{k} \cdot \mathbf{r}} \longrightarrow (2\pi)^3 \delta^{3}(\mathbf{k}_p-\mathbf{k}_s -\mathbf{k}_i).
\label{infinito}
\end{equation}
Considering steady state fields, on the other hand, the time-integral in (\ref{Phi}) can be also reduced, one has
\begin{equation} \int_{t_0}^t  d\tau  e^{i  (\omega_s+\omega_i-\omega_p)\tau}  \longrightarrow  \int^\infty_{-\infty}  d\tau  e^{i  (\omega_s+\omega_i-\omega_p)\tau}= 2\pi \delta(\omega_p-\omega_s-\omega_i).
\label{steady}
\end{equation}
Notice that the expressions at the right in (\ref{infinito}) and (\ref{steady}) correspond to the conservation of the momentum (\ref{momentum}) and energy (\ref{energy}) respectively. Introducing these last results into (\ref{Phi}) we arrive at the simplified expression
\begin{equation}
\Phi(\mathbf{k_i},\omega_i,\mathbf{k_s}, \omega_s)=\eta \alpha(\mathbf{k}_s +\mathbf{k}_i,\omega_s + \omega_i),
\label{Phi2}
\end{equation}
with $\eta$ a global factor that will be dropped in the sequel. Now, considering the phase-matching condition defined by Equations (\ref{energy}) and (\ref{momentum}), let us assume that the $\alpha$-function can be factorized as the product of a function $E$ depending only on the wave vector information and a function $F$ that only depends on the frequencies, we can write
\be
\Phi(\mathbf{k_i},\omega_i,\mathbf{k_s}, \omega_s)=\eta E (\mathbf{k}_s +\mathbf{k}_i) F(\omega_s + \omega_i).
\label{factor}
\ee
At this stage, it is convenient to separate the wave vectors $\mathbf k_r$ into their transverse $\mathbf q_r$ and longitudinal $k_r$ components:
\be
\mathbf k_r = \mathbf q_r + k_{rz} \mathbf e_{rz}, \quad \mathbf q_r= q_{rx} \mathbf e_{rx} + q_{ry} \mathbf e_{ry}, \quad \vert \vert \mathbf  e_{r\ell} \vert \vert = 1, \quad r=p,s,i, \quad \ell =x,y,z.
\label{compo}
\ee
Using paraxial beams \cite{Sal91}, the wave vectors (\ref{compo}) are reduced to their transverse components \cite{Mol05} and, assuming that the factor functions in (\ref{factor}) are Gaussian distributions (see e.g. \cite{Joo96}),
\be
E(\mathbf A) = \exp ( -\sigma^2 \vert \vert \mathbf A \vert \vert^2), \quad F(\omega) = \exp(-\delta^2 \omega^2), 
\label{gauss}
\ee
we arrive at the expression
\begin{equation}
\Phi(\mathbf{q_i},\omega_i,\mathbf{q_s}, \omega_s)=\eta \exp ( -\sigma^2 \vert \vert \mathbf q_s + \mathbf q_i \vert \vert^2 ) F(\omega_s + \omega_i).
\end{equation}
Let us associate the spatial and spectral sensitivity of the detectors to the Gaussian-profile filters
\be
{\cal F}_{S} (\mathbf q_r) = \exp\left( -\vert \vert \vec{\sigma} \cdot \mathbf q_r \vert\vert^2\right), \quad \mbox{and} \quad {\cal F}_F(\omega_r) = e^{-\alpha_r^2 \omega_r^2},
\label{filters}
\ee
respectively. Here the components of $\vec \sigma =(\sigma_{rx}, \sigma_{ry})$, and $\alpha_r$, correspond to spatial and frequency collection modes. Hence, we finally write
\begin{equation}
\Phi(\mathbf{q_i},\omega_i,\mathbf{q_s}, \omega_s)=\Phi_{\omega} (\omega_i,\omega_s)\Phi_{\mathbf q} (\mathbf{q_i},\mathbf{q_s})
\label{fifin}
\end{equation}
where the functions
\be
\begin{array}{c}
\Phi_{\omega}(\omega_i, \omega_s)= \eta F_p(\omega_s+ \omega_i) \mathcal{F}_F(\omega_i)  \mathcal{F}_F (\omega_s),\\[2ex]
\Phi_{\mathbf{q}} (\mathbf{q_i},\mathbf{q_s})= \exp ( -\sigma^2 \vert \vert \mathbf q_s + \mathbf q_i \vert \vert^2 ) \mathcal{F}_S (\mathbf{q}_s) \mathcal{F}_S (\mathbf{q}_i),
\end{array}
\ee
encode information on the time and spatial correlations of the biphoton states. We are interested in the spatial correlations as they must be observed in the lab, so that it is convenient to make a Fourier transformation from the $\mathbf q_r =(q_{rx}, q_{ry})$ planes into the Cartesian ones $\mathbf r_r = (x_r, y_r)$, $r=i,s$,  as follows
\begin{equation}
\widetilde \Phi(\mathbf{r_i},\mathbf{r_s})\equiv \widetilde \Phi( x_i,y_i,x_s,y_s)
= N \int d^2 r_i\int d^2 r_s  \Phi_{\mathbf{q}}(\mathbf{q_i},\mathbf{q_s}) e^{-i \mathbf{q}_s \cdot \mathbf{r}_s }e^{-i \mathbf{q}_i \cdot \mathbf{r}_i },
\label{resul}
\end{equation}
with $N$ a normalization constant.

\subsection{Main results}
\label{main}

The function $\widetilde \Phi(\mathbf{r_i},\mathbf{r_s})$, derived in Equation~(\ref{resul}) of the previous section, corresponds to the quantum theoretical predictions for the distribution rates of coincidences in the counting of photons at the detection zone. The idler and signal channels are measured by photo-collectors located at the positions $(x_i, y_i)$ and $(x_s, y_s)$ on the ring described in Figure~\ref{fig1}. Diverse combinations of the cartesian variables $(\mathbf{r_i},\mathbf{r_s})$ in the quantum approach predictions presented here have been depicted in the top line of Figure~\ref{fig2}. For comparison, we have included also the predictions obtained by using the geometric model \cite{Pro09,Pro10} (middle line) and the corresponding experimental data (bottom line) as well.  The collecting of photons in the lab was performed by displacing the idler and/or signal detectors according to the vectors $\mathbf r_i$ and $\mathbf r_s$ described in the figure caption. Some of the main characteristics of the experimental setup are included in Section~\ref{experiment}, more details can be found in \cite{Pro09}. The agreement between the results obtained from these three different approaches is good enough as this can be appreciated from the global behavior of the distributions in Figure~\ref{fig2}. In all cases, horizontal cuts produce ellipse-like contours that are oriented at $45^o$ in the horizontal-horizontal ($x_i x_s$) and vertical-vertical ($y_i y_s$), at $90^o$ in the horizontal-vertical ($x_s y_i$), and at $0^o$ for the vertical-horizontal ($y_sx_i$) positions of the photodetectors. The projections of the distribution $\widetilde \Phi(\mathbf{r_i},\mathbf{r_s})$ into the above indicated planes are shown in Figure~\ref{fig3}. The ellipticity of the coincidence rates is clearly manifest in all cases, some differences can be found in the size of the semi-axes because the involved units (in all cases we use arbitrary units). The advantage of the theoretical model discussed here is that the inclusion of filters like the ones defined in (\ref{gauss}) allows the fixing of the semi-axes size accordingly. Our approach lies on the information of the pump beam that is encoded in the parameter $\delta$ that defines the width of the SPDC created photon pairs ring in the detection zone of Figure~\ref{fig1}. As we have indicated in the introduction of the paper, this parameter determines the geometric profile of the distribution rates of spatial coincidences illustrated in figures~\ref{fig2} and \ref{fig3}.

\begin{figure}[ht]
\centering
\includegraphics[scale=0.25]{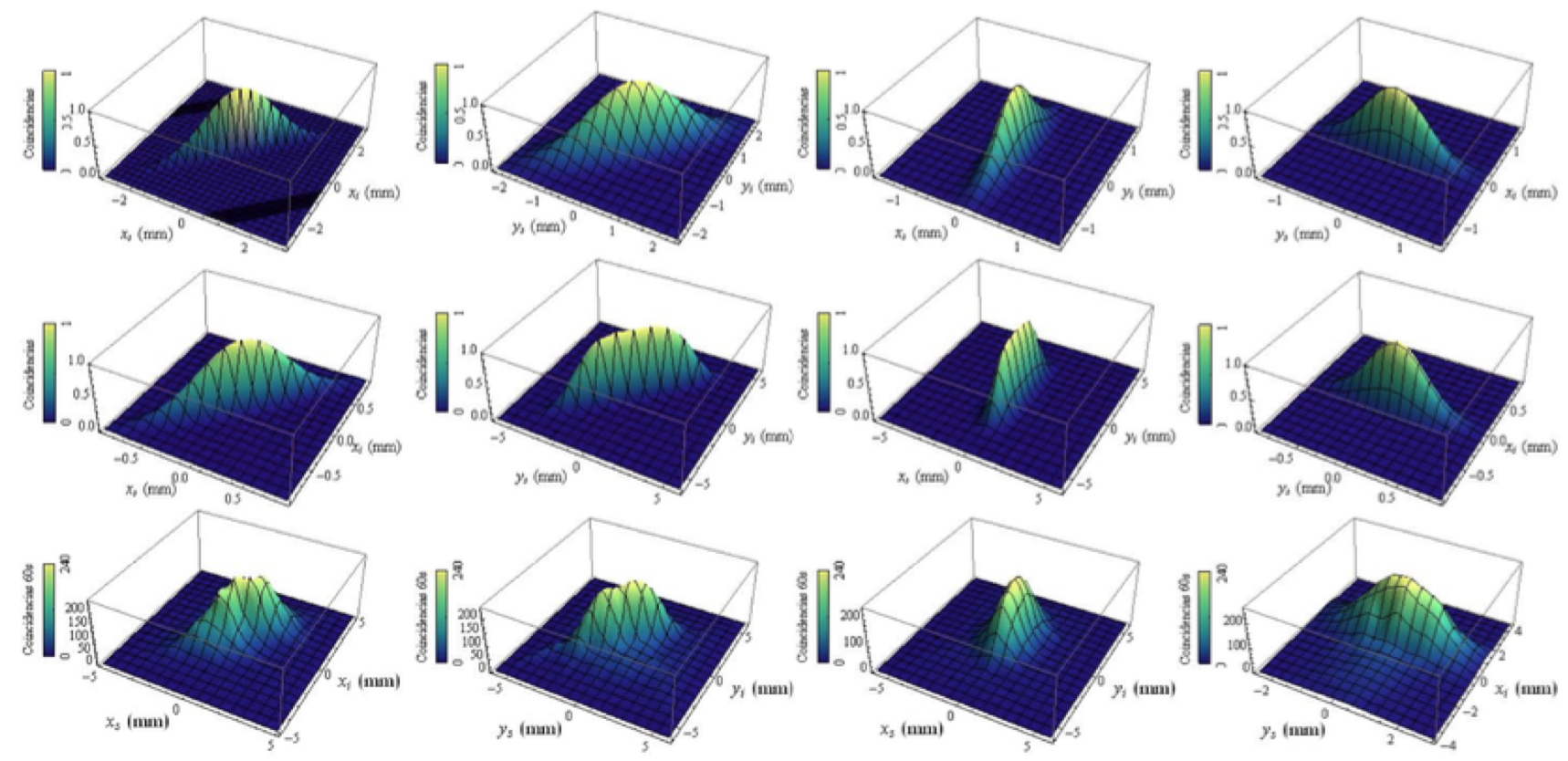}  
\caption{\label{fig2} \footnotesize
Distribution rates of spatial coincidences (\ref{resul}) calculated by using the quantum theory of this work (top) and the geometric model reported in \cite{Pro09,Pro10} (middle). The corresponding experimental observations \cite{Pro09} are included at the bottom line.  From left to right they correspond to $\widetilde \Phi (x_i,y_i,x_s,y_s)$ with $y_i=y_s=0$, $x_i=x_s=0$, $x_i=y_s=0$ and $y_i=x_s=0$.
}
\end{figure}

\begin{figure}[h]
\centering
\includegraphics[scale=0.25]{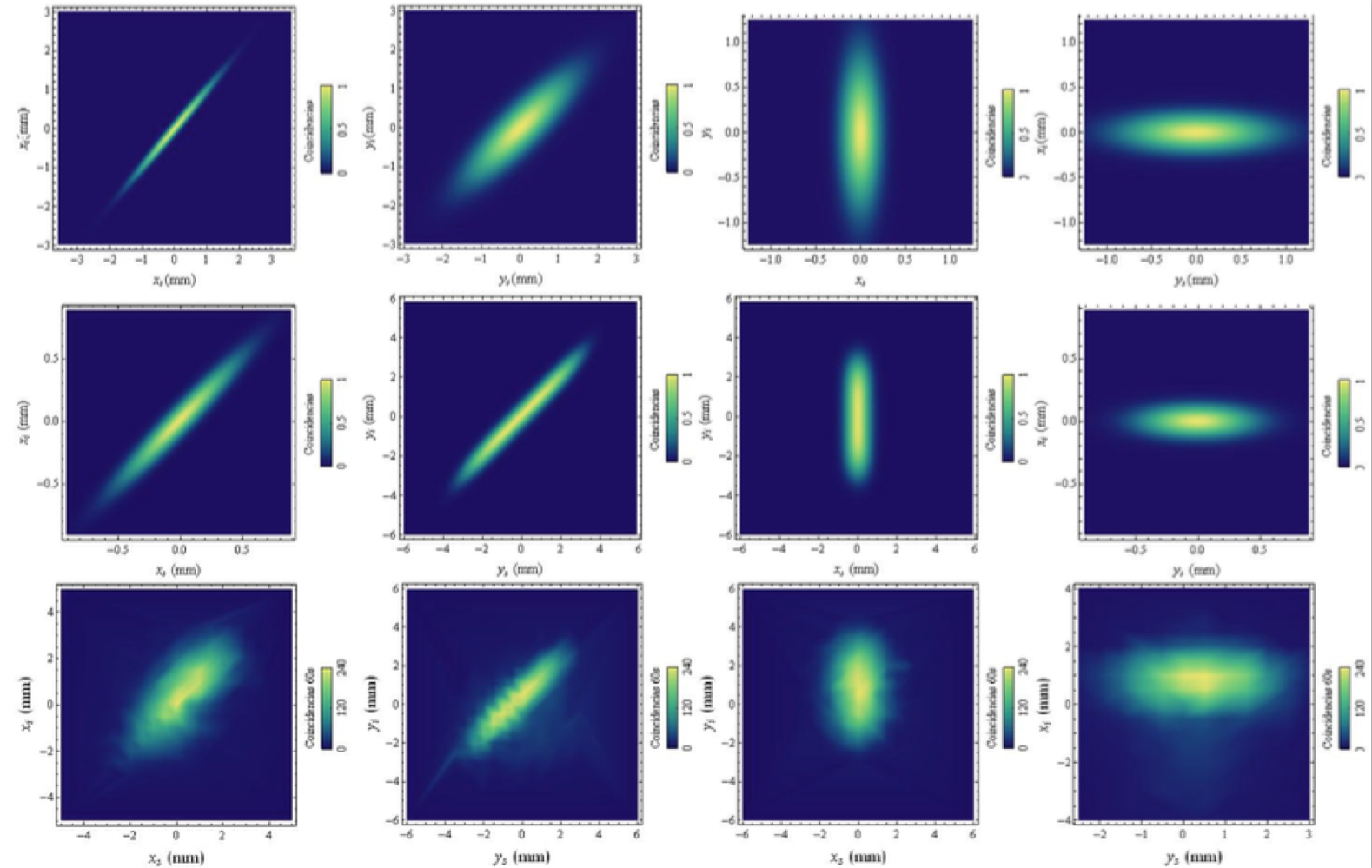}  
\caption{\label{fig3} \footnotesize
Projections of the distribution rates depicted in Figure~\ref{fig2} onto the cartesian planes $(x_s, x_i)$, $(y_s, y_i)$, $(x_s, y_i)$ and $(y_s, x_i)$.
}
\end{figure}

\subsubsection{Experimental setup \cite{Pro09,Pro10}.}
\label{experiment}

We used a BBO-I crystal with effective length $L_z = 2$ mm, and a cross area of 5 mm $\times$ 5 mm. The angle between the optical axis of the crystal and the pump direction ($z$-coordinate) was $\theta_{pc} = 30^o$. The crystal was pumped by a 100 mW violet laser diode operating at 405.38 nm and bandwidth 0.78 nm. The idler (signal) detector $D_i$ ($D_s$) was located $L_D = 0.849$ m from the center of the crystal and was free to move in the plane transverse to $\mathbf k_i (\mathbf k_s)$. We fixed $\varphi_i = 85.98$ mrad and $\varphi_s = 80.09$ mrad by adjusting the position of $D_i$ and $D_s$ in the $xz$-plane for maximum counting rate. Thus, phase-matching was attained with $D_i$ ($D_s$) far from the pump beam $d_i = 7.3$ cm ($d_s = 6.8$ cm) and a degenerated wavelength $\lambda_{i,s} = 810$ nm. The remaining part of the pump beam was blocked to avoid perturbations in the measuring zone. Before arriving at the coincidence counter, each of the correlated photons passed trough a 400$\mu$m pinhole and a $810 \pm 10$ nm filter, this last reducing the irradiance of the residual pump light as well as the background noise. Then, the pairs were collected with 5 mm spherical lenses, connected to avalanche photodiodes by optic fibers. The interval of coincidences was of 10 ns, with a counting time of 60 s. The counting rates of each channel as well as the coincidence rate were obtained by displacing the photon collectors in the $x_{i,s}$ and $y_{i,s}$ directions.

\section{Conclusions}

We have developed an approximate expression for the distribution rate of spatial coincidences in the detecting of twin photons generated by spontaneous parametric down conversion. We were guided by a geometric model that was already verified in the lab \cite{Pro09,Pro10}. The main point of such a model is that the spatial distribution of the SPDC created photons can be considered as having a normal (Gaussian) profile because the predicted spatial region of detection is a ring of width $2\delta$ that is centered at the pump beam. Thus, the pairs of created photons will be find with certainty in a point on the ring, one in the antipodal position of the other. The parameter $\delta$ that defines the width of the ring is a function of the general properties of the pump beam. In the geometric model it is assumed that the pump beam is paraxial with a negligible transverse  width and having a very small spectral width (i.e., in a first approach the pump beam is considered as having no structure). More elaborated approximations can be developed by taking into account the spatial shape of the pump bean \cite{Joo96,Mol05}. For instance, the pump wave vector occupies a cone of finite light if the pump beam has a finite transverse width, so that the conservation of momentum can be satisfied in more than one way \cite{Joo96}. The transverse width influences also the ellipticity of the spatial distributions, specially for highly focused beams \cite{Mol05}. Our approach can be extended to include more specific information of the pump beam by making $\delta$ a function of the corresponding parameters. However, the global elliptical form of the distribution rates must be preserved, as this effect has been demonstrated experimentally. 

\subsection*{Acknowledgments}

The support of CONACyT project 152574 and IPN project SIP-SNIC-2011/04 is acknowledged. 


\end{document}